\documentclass[manuscript]{aastex} 
\newcommand{\etal}{{\it et al.\/}}

\begin{document}
\pagenumbering{arabic}

\title{HOW DID GLOBULAR CLUSTERS FORM?}
\author{Sidney van den Bergh}
\affil{Dominion Astrophysical Observatory, Herzberg Institute of Astrophysics, 
National Research Council of Canada, 5071 West Saanich Road, Victoria, BC, 
Canada  V9E 2E7}
\email{sidney.vandenbergh@nrc.ca}

\begin{abstract}

It is suggested that there have been at least two physically distinct epochs of 
massive cluster formation. The first generation of globular clusters may have 
formed as halo gas was compressed by shocks driven inward by ionization fronts 
generated during cosmic reionization \citep{cen2001} 
at $z \sim 6$ \citep{beckeretal2001}. On the 
other hand a second generation of massive clusters might have been formed at 
later times by compression, and subsequent collapse, of giant molecular clouds. 
In some cases this compression may have been triggered by heating of the 
interstellar medium via collisions between gas-rich disk galaxies. 
It is also suggested that the present specific globular cluster 
frequency in galaxies was mainly determined by the peak rate of 
star creation, with elevated peak rates of star formation per unit area 
resulting in high present specific globular cluster frequencies
\citep{larsenandrichtler2000}. 

\end{abstract}


\section{INTRODUCTION}

Twenty years ago Bill Harris and I \citep{harrisandvandenbergh1981} introduced 
the specific globular cluster frequency S, i.e the number of  clusters per unit 
of parent galaxy luminosity, which was defined as
\\
\hspace*{6cm}{S = N x $10^{0.4 (M_{v} + 15)}$ }.\hspace*{6cm}(1)  
         
Observations of S in a wide variety of parent galaxy types \citep{harris1991} 
showed large variations in the galaxy-to-galaxy specific globular cluster 
frequency. Major trends were that: (1) S is generally found to be higher in 
elliptical galaxies than it is for disk galaxies, (2) Many (but not all) central 
cluster galaxies in galaxy clusters have above-average S values, (3) among 
ellipticals S does not appear to depend strongly on parent galaxy luminosity. 
Notable peculiarities are that (a) M32 contains no globulars [even though 10-20 
are expected (van den Bergh 2000, p. 168)], and (2) the Fornax dwarf spheroidalsystem 
contains an anomalously high number of globulars. It is the purpose of the 
present note to look in to some of the factors that might have affected the 
presently observed S values of galaxies.

\section{CLUSTER FORMATION DURING MERGERS}

\citet{vandenbergh1979} hypothesized that the formation of massive star clusters 
might have been triggered by strong shocks. Such shocks are expected to be 
common in starburst galaxies and to be rare in quiescent dwarf galaxies such as 
IC 1613. Alternatively \citet{ashmanzepf2001} have recently suggested that the 
formation of massive clusters is favored by the high pressure in the 
interstellar medium (ISM) that is expected to prevail in merging spirals and in 
starburst galaxies. Ashman \& Zepf argue that a high-pressure ISM will compress 
giant molecular clouds and trigger star and cluster formation within them. 
Perhaps bound clusters are only produced from the compact cores in such clouds, 
whereas unbound associations might be formed from their lower density envelopes. 
Ashman \& Zepf propose that the formation of massive clusters by the compression 
of giant molecular clouds might account for the existence of numerous massive 
intermediate-age clusters in colliding galaxies such as NGC 4038/4039 
\citep{whitmoreetal1999}. Because the frequency of such mergers 
\citep{carlbergetal2000} decreases with time, the massive clusters produced by 
such mergers should be most common at large redshifts.

\section {CLUSTER FORMATION AFTER REIONIZATION}

An alternative scenario for the formation of massive clusters has recently been 
proposed by \citet{cen2001}.  He points out that reionization of the Universe 
\citep{gnedinlahavandrees2001,beckeretal2001} 
will drive ionization fronts into the gaseous halos of 
protogalaxies, resulting in convergent shocks. These shocks will compress the 
resident gas by factors of $\sim$ 100. Cen argues that this compression will 
lead to fragmentation and to the formation of clusters with masses in the range 
$10^3 - 10^6$ M$_{\odot}$. The clusters so formed might be identified with the 
first generation of globular clusters. The separation in time between this first 
generation of globular clusters and later star clusters is particularly clean in 
the Large Magellanic Cloud \citep{suntzeff1992} where all 13 globulars have ages 
$>$10 Gyr, whereas all but one of the other massive clusters have ages $\leq$3 
Gyr. In our own Milky Way system the situation appears to be more complex, with 
a generation of halo globular clusters that have below-average ages and masses 
having formed in the outer halo beyond R$_{gc}$ = 15 kpc 
\citep{vandenbergh1998}. Possibly these outer halo clusters were originally 
formed in dwarf spheroidal galaxies that were later captured (and tidally 
destroyed) by the Galaxy. Many of such dwarf spheroidals 
(van den Bergh 2000, pp. 243 - 264) are known to have 
undergone major bursts of star formation (during which massive clusters might 
have formed) a few Gyr after their formation.

\section {THE SPECIFIC CLUSTER FREQUENCY}

A possible key to understanding the galaxy-to-galaxy variations in the specific 
globular cluster frequency S is provided by \citet{larsenandrichtler2000}, who 
have determined the fraction on the ultraviolet light of various galaxies that 
is emitted by massive [M $>$ 4000 M$_{\odot}$] and young [age $<$ 0.5 Gyr] 
clusters. These authors find that the fraction of the UV light of galaxies that 
is emitted by such young massive clusters ranges from $\sim 15\%$ in the 
starburst galaxy NGC 3256 to $\sim 0\%$ in the quiescent dwarf IC 1613. The low 
rate of cluster formation in IC 1613 was first noted by \citet{baade1963}, and 
was subsequently confirmed by \citet{hodge1978} and \citet{vandenbergh1979}. 
Figure 1 shows a plot, based on Larsen \& Richtler's data, of the relation 
between the percentage of the total U light of a galaxy that is emitted by 
massive young clusters and the mean rate of star formation per unit area. These 
data show that the fraction of the U light of galaxies that is generated by 
young clusters is proportional to the rate of star formation  per unit area to 
the power $\sim$ 1.2. This result suggests that the present specific cluster 
frequency in galaxies will, to a large extent, be determined by the peak rate of 
star formation per unit area during their evolutionary history. The high S 
values observed in some central galaxies of rich clusters might therefore be due 
to a high surface density of star formation early in their history. By the same 
token the observation that elliptical galaxies have higher mean S values than 
spirals may be interpreted to mean that the peak rate of star formation in E 
galaxies was higher than that in disk galaxies. This speculation is supported by 
the observation that the ratio of alpha elements(producedby short-lived SNe II) 
to iron (formed in longer-lasting SNe Ia) is greater in ellipticals than it is 
in spiral disks \citep{wheelersnedentruran1989}.

\section {SUMMARY AND CONCLUSIONS}

It is suggested that there are two families of massive star clusters. The first 
of these (the classical globular clusters) were produced at high redshift when 
reionization sent ionization fronts crashing into the gaseous halos of 
protogalaxies. The formation of the second class of massive clusters was 
triggered by the squeezing of giant molecular clouds by the hot ISM in which 
they were embedded, by starbursts initiated by collisions between gas-rich 
galaxies. The fraction of all star formation that ends up in the form of massive 
bound star clusters depends critically on the star formation rate per unit area.  

It is a pleasure to thank Soeren Larsen and Steve Zepf for their helpful 
comments on an earlier version of this paper. I am also indebted to the 
referee for a number of helpful suggestions.

\clearpage

\centerline{Figure Caption}

Figure 1. Percentage of ultraviolet light generated by massive young clusters 
versus rate of star formation in M$_{\odot}$ yr$^{-1}$ kpc$^{-2}$.  The slanted 
line shows that the fraction of a galaxy's light emitted by massive young 
clusters is approximately proportional to the star formation rate per unit are 
to the power 1.2.

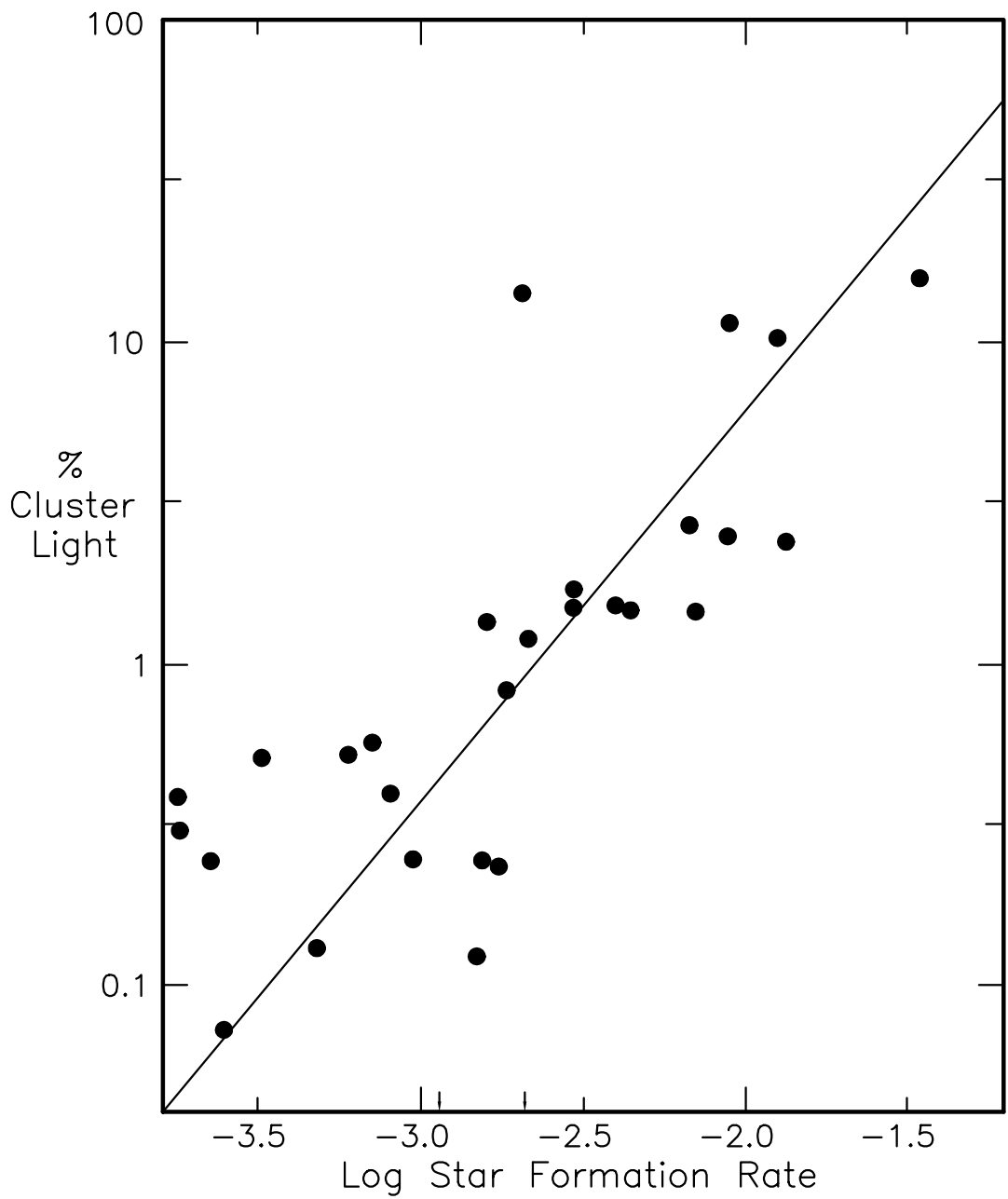
\end{document}